\documentstyle[epsfig,aps,multicol]{revtex}
\begin{document}
\title{New Tetrahedral Global Minimum for the 98-atom Lennard-Jones Cluster} 
\draft
\author{Robert H. Leary}
\address{San Diego Supercomputer Center, University of California, San Diego, 
San Diego, CA 92138, USA}
\author{Jonathan P. K. Doye}
\address{University Chemical Laboratory, Lensfield Road, Cambridge CB2 IEW, 
United Kingdom}
\date{\today}
\maketitle
\begin{abstract}
A new atomic cluster structure corresponding to the global minimum of
the 98-atom Lennard-Jones cluster has been found using 
a variant of the basin-hopping global optimization algorithm.
The new structure has an unusual tetrahedral symmetry with an energy of 
$-543.665361\epsilon$, which is $0.022404\epsilon$ lower than the previous putative 
global minimum. The new LJ$_{98}$ structure is of
particular interest because its tetrahedral symmetry establishes it as one of
only three types of  exceptions to the general pattern of icosahedral structural
motifs for optimal LJ microclusters.
Similar to the other exceptions the global minimum is difficult to find because
it is at the bottom of a narrow funnel which only becomes thermodynamically 
most stable at low temperature.
\end{abstract}

\begin{multicols}{2}

The determination of the global minima of Lennard-Jones (LJ) clusters
by numerical global optimization techniques has been been intensely studied 
in the size range $N$=13-147 by both chemical physicists and 
applied mathematicians\cite{WalesS99,Wille99}. 
The LJ potential, which is given by
\begin{equation}
E = 4\epsilon \sum_{i<j}\left[ \left(\sigma\over r_{ij}\right)^{12} - \left
(\sigma\over r_{ij}\right)^{6}\right],
\end{equation}
where $\epsilon$ is the pair well depth and $2^{1/6}\sigma$ is the
equilibrium pair separation, 
is a simple yet reasonably accurate model of the interactions between
heavy rare gas atoms.
In general, there has been good agreement between physical
measurements on rare gas clusters from electron diffractometry \cite{Farges86} 
and mass spectrometry \cite{Echt81,Harris84}
and computational global optimization results regarding magic number sizes
and corresponding cluster geometries \cite{Northby87}.  
Both approaches find that Mackay icosahedra \cite{Mackay} are the dominant structural motif.

The LJ microcluster problem has also become a benchmark for evaluating
global optimization algorithms.  The number of local minima (excluding 
permutational isomers) on the potential energy surface (PES) is believed to 
grow exponentially with $N$ \cite{Tsai93a,Still99} 
and is estimated to be of the order of $10^{40}$ for $N$=98.
A wide variety of global optimization techniques including simulated 
annealing \cite{Wille87}, genetic algorithms \cite{Niesse96a,Deaven96,Wolf98}, 
smoothing and hypersurface deformation techniques \cite{Kostrowicki,Pillardy}, 
lattice methods \cite {Northby87,Xue94}, growth sequence
analysis \cite {HoareP71,Leary97}, and tunneling \cite{Barron96} have been
applied to the problem.  
Unbiased methods that make no assumptions regarding cluster geometry are of the
most interest, since these have the best chance of successful generalization
to more complex potentials such as those in the protein folding problem. 

\begin{center}
\begin{figure}
\epsfig{figure=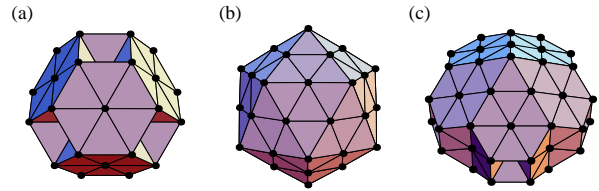,width=8.2cm}
\begin{minipage}{8.5cm}
\caption{\label{fig:structures} Three particularly stable examples of the known
morphologies for LJ clusters:
(a) the 38-atom fcc truncated octahedron, (b) the 55-atom Mackay icosahedron, 
and (c) the 75-atom Marks decahedron.
}
\end{minipage}
\end{figure}
\end{center}

Most of the global minima in this size range were first found by Northby
in a lattice-based search of icosahedral structures \cite{Northby87}.
These structures consist of a core Mackay icosahedron (Figure \ref{fig:structures}b)
surrounded by a partially filled outer shell.  
More recently, there have been a number of improvements in 
some of these putative global minima. 
Firstly, further refinements to Northby's algorithm, particularly the relaxation
of the assumption that the core Mackay icosahedron is always complete, 
has a led to a number of new global minima \cite{Xue94,Leary97,Coleman,Barron97}.
Secondly, consideration of particularly stable face-centred-cubic (fcc) and 
decahedral forms has also led to new global minima \cite{Barron96,Doye95c,Doye95d}. 
At $N$=38 the global minimum is a fcc truncated octahedron 
(Figure \ref{fig:structures}a) and at $N$=75-77 and 102-104
the global minima are based on Marks decahedra (Figure \ref{fig:structures}c).
Thirdly, powerful unbiased global optimization algorithms, 
particularly the basin-hopping \cite{WalesD97} and 
genetic algorithms\cite{Deaven96,Wolf98}, have recently begun to catch up 
with those methods that incorporate particular physical insights
into the LJ problem, and are now able to find all the known lowest-energy minima.

\begin{center}
\begin{figure}
\epsfig{figure=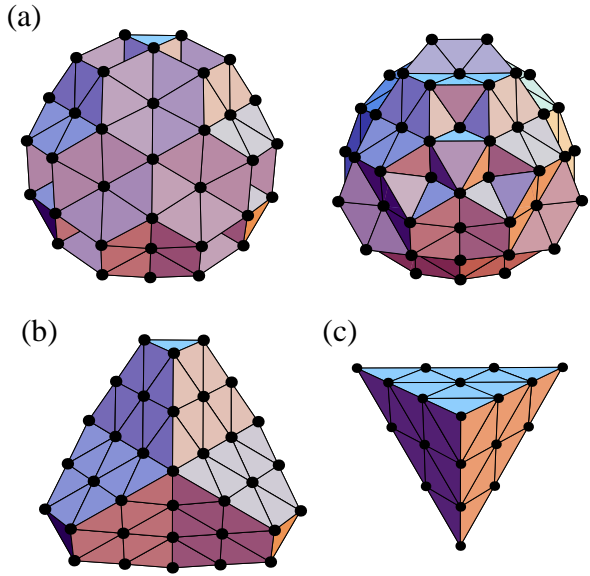,width=8.2cm}
\vglue 2mm
\begin{minipage}{8.5cm}
\caption{\label{fig:LJ98} 
(a) Front and back views of the new LJ$_{98}$ global minimum. 
(b) The 56-atom stellated tetrahedron
and (c) the 20-atom tetrahedron that are at the centre of this structure.
}
\end{minipage}
\end{figure}
\end{center}

Given this combined attack on the LJ optimization problem, it might have been
imagined that all the global minima for $N<150$ had been found. 
Here, however, we report a new lowest-energy structure for LJ$_{98}$,
which has an energy of $-543.665361\epsilon$ and $T_d$ point group 
symmetry. This compares to an energy of $-543.642957\epsilon$ 
for the previous icosahedral putative global minimum which was found by 
Deaven {\it et al} \cite{Deaven96}. 
The LJ$_{98}$ global minimum is organized around 
a central fcc tetrahedron with four atoms on each edge (Figure \ref{fig:LJ98}c).  
Four additional fcc tetrahedrons (minus apices) are erected 
over the faces of the central tetrahedron to form a 56-atom stellated
tetrahedron (Figure \ref{fig:LJ98}b).  
An additional 42 atoms decorate the closed-packed sites on the surface of the stellated
tetrahedron to complete the structure (Figure \ref{fig:LJ98}a).    
The new LJ$_{98}$ structure is of particular interest because its 
tetrahedral symmetry establishes it as only the third known type of 
exception to the general pattern of icosahedral structural
motifs for optimal LJ microclusters, and the first to be discovered by
an unbiased optimization method.

Given its unusual structure one might wonder why it is so low in energy.
For LJ clusters optimizing the energy is a balance between maximizing the 
number of nearest neighbours and minimizing the strain energy (the energetic
penalty for nearest-neighbour distances deviating from the
equilibrium pair value) \cite{Doye95c}.
The spherical shape and high proportion of $\{111\}$ faces gives the structure
a large number of nearest neighbours (432 compared to 437 for the lowest-energy
icosahedral minimum and 428 for the lowest-energy decahedral structure), 
whilst its strain energy is intermediate between icosahedral and decahedral structures. 
The lower strain energy allows it to be lower in energy than the icosahedral minima,
even though it has fewer nearest neighbours.
The strain in the structure is focussed around the six edges of the 
central fcc tetrahedron. The atoms along these edges have the same local coordination
as atoms along the five-fold axis of a decahedron.

It is also natural to ask how general this structure is. Firstly, analogous
structures can be formed with smaller and larger tetrahedra at their core. The
previous one in this series is at $N$=34 and the next one is at $N$=195. 
However, these structures are not energetically competitive: the former
because it has too high a proportion of $\{100\}$ faces, and the latter
because it is not sufficiently spherical. Secondly, the structures of the
other non-icosahedral LJ global minima have been experimentally observed
for gold\cite{Cleveland97b} and nickel\cite{Parks97} clusters, and found to be
particularly stable in theoretical calculations of transition
metal clusters \cite{Doye98c}. 
Therefore, we performed some optimization calculations for the Sutton-Chen
family of potentials \cite{Sutton90}. The tetrahedral
structure was lowest in energy for silver, but a decahedral minimum was lower
in energy for nickel and a fcc minimum for gold.
This is consistent with previous results, which indicted that, of these three metals,
silver clusters exhibited ordered structures with the most strain\cite{Doye98c}.

The new LJ$_{98}$ optimum was found using a variant of the 
basin-hopping global optimization algorithm \cite{WalesD97}.  
The key idea behind the algorithm is the mapping of the 
original LJ potential energy function, $E({\bf x})$, for each point {\bf x} 
on the $3N$-dimensional Cartesian coordinate space onto a ``transformed'' 
energy function, $T({\bf x})$. $T({\bf x})$ takes the value of $E({\bf x})$
at the local minium, ${\bf x_{min}}$, arrived at by applying a given local 
optimization procedure, such as the conjugate gradient algorithm, 
with ${\bf x}$ as the starting point for the algorithm.  
Thus $T({\bf x})$ is a ``plateau'' function that takes on the constant value 
$E({\bf x_{min}})$ on the catchment basin surrounding each local minimum
${\bf x_{min}}$.  $T({\bf x})$ is a lower bound to $E({\bf x})$ and coincides 
with $E({\bf x})$ at all of the latter's local minima, 
but all barriers are removed in the $T({\bf x})$ landscape
and transitions bewteen catchment basins can take place all along the
basin boundaries. 

The original basin-hopping algorithm consists of a Metropolis search of the
transformed landscape, $T({\bf x})$, using a Monte Carlo sampling procedure
to move between local minima.  In the variant used in the discovery of 
LJ$_{98}$ \cite{Learyinprep}, the Metropolis criterion of accepting uphill moves with a 
probability that is an exponentially decreasing function of the energy 
increment is abandoned in favor of only accepting downhill moves.  
The algorithm is restarted from a fresh random starting local minimum whenever 
progress stalls for a sufficiently large number of move attempts. 
The variant was successful in locating the LJ$_{98}$ global minimum in 6 of 1000 random starts, 
with a mean computational time between encounters of about 30 hours on 
a 333 MHz Sun Ultra II processor.  
This structure has also been subsequently found using the original
basin-hopping algorithm \cite{Walespersonal}.

These results show that the LJ$_{98}$ global minimum is particularly difficult to find. 
The origins of this difficulty are probably similar to the other
non-icosahedral clusters. Analyses of the PESs of LJ$_{38}$ and LJ$_{75}$ 
using disconnectivity graphs have shown that they consist of a 
wide icosahedral ``funnel'' \cite{Leopold,Bryngelson95} 
and a much narrower funnel leading to the global 
minimum \cite{Doye99c,Doye99f}.
On relaxation down the PES the cluster is much more likely to enter the icosahedral funnel,
where it is then trapped because of the large (free) energy barriers to escape
from this funnel into the funnel of the global minimum. 

This situation is compounded by the thermodynamics of these clusters \cite{Doye98a,Doye98e}. 
The icosahedral funnel has a larger entropy because of the larger number of 
low-energy minima, and so the funnel of the global minimum is
only lowest in free energy at low temperatures. 
Therefore, at temperatures where the dynamics occur at a reasonable rate there is 
a thermodynamic driving force to enter the icosahedral funnel.
For LJ$_{98}$ there are at least 114 minima that are lower in energy than the 
second lowest-energy minimum in the tetrahedral funnel,
and so the global minimum is only lowest in free energy below $T$=0.0035$\epsilon k^{-1}$ 
(a typical melting temperature for a LJ cluster is 0.3$\epsilon k^{-1}$).
This transition temperature is markedly lower than for 
LJ$_{38}$ \cite{Doye99f} or LJ$_{75}$ \cite{WalesD97}.

The basin-hopping transformation of the PES helps to ameliorate 
some of these difficulties. The transformation changes the thermodynamics
so that the global minimum still has a significant occupation probability at 
temperatures where the cluster can escape from the icosahedral funnel.
However, on relaxation down the PES the system is still much more likely 
to enter the icosahedral funnel. 
For example, our optimization runs were fifteen times more likely to terminate at 
the lowest-energy LJ$_{98}$ icosahedral minimum than at the global minimum.

Coordinate files for the new LJ$_{98}$ structure, as well as all other 
putative LJ microcluster global optima, can be found in the Cambridge 
Cluster Database \cite{Web}.

\end{multicols}
\end{document}